\documentclass[10pt]{article}
\usepackage{fullpage}
\usepackage{amsmath}
\usepackage{amssymb}
\usepackage[dvips]{epsfig}
\usepackage{color}

\def\##1{\underline{#1}}
\def\=#1{\underline{\underline{#1}}}

\def\+#1{\underline{\bf #1}}
\def\*#1{\underline{\underline{\bf #1}}}

\def\r#1{(\ref{#1})}
\def\l#1{\label{#1}}
\def\c#1{\cite{#1}}

\def\le{\left(}
\def\ri{\right)}
\def\les{\left[}
\def\ris{\right]}
\def\lec{\left\{}
\def\ric{\right\}}

\def\.{\cdot}

\def\epso{\epsilon_{\scriptscriptstyle 0}}

\def\muo{\mu_{\scriptscriptstyle 0}}

\def\eps{\epsilon}

\begin{document}

\begin{center}

{\bf {\LARGE A simple construction for a cylindrical  \vspace{6pt} \\
 cloak via inverse homogenization}}

\vspace{10mm} \large

 Tom H. Anderson\footnote{E--mail: T.H.Anderson@sms.ed.ac.uk}\\
{\em School of Mathematics and
   Maxwell Institute for Mathematical Sciences\\
University of Edinburgh, Edinburgh EH9 3JZ, UK}\\
 \vspace{3mm}
 Tom G. Mackay\footnote{E--mail: T.Mackay@ed.ac.uk}\\
{\em School of Mathematics and
   Maxwell Institute for Mathematical Sciences\\
University of Edinburgh, Edinburgh EH9 3JZ, UK}\\
and\\
 {\em NanoMM~---~Nanoengineered Metamaterials Group\\ Department of Engineering Science and Mechanics\\
Pennsylvania State University, University Park, PA 16802--6812,
USA}\\
 \vspace{3mm}
 Akhlesh  Lakhtakia\footnote{E--mail: akhlesh@psu.edu}\\
 {\em NanoMM~---~Nanoengineered Metamaterials Group\\ Department of Engineering Science and Mechanics\\
Pennsylvania State University, University Park, PA 16802--6812,
USA}\\
and\\
 {\em Materials Research Institute\\
Pennsylvania State University, University Park, PA 16802, USA}

\end{center}

\vspace{4mm}

\normalsize

\begin{abstract}

An effective  cylindrical cloak may be conceptualized as
 an assembly of adjacent local neighbourhoods, each of which is made from  a
   homogenized composite material (HCM). The HCM is required to be a certain uniaxial
   dielectric-magnetic material, characterized by positive-definite constitutive
   dyadics. It can arise from the homogenization of remarkably
   simple component materials, such
as  two isotropic dielectric-magnetic materials, randomly
distributed as oriented spheroidal particles. By carefully
controlling the spheroidal shape of the component particles, a high
degree of HCM anisotropy may be achieved, which is necessary for the
cloaking effect to be realized.
 The inverse Bruggeman
formalism can provide estimates of the   shape and constitutive
parameters for the  component materials, as well as their volume
fractions.

\end{abstract}


\vspace{5mm} \noindent  {\bf Keywords:} cloak, geometric optics,
Bruggeman formalism, inverse homogenization

\section{Introduction}

The problem of light propagation in a homogeneous material occupying
a region of complex shape  can be reversibly transformed into the
problem of light propagation in an inhomogeneous material occupying
a region of simple shape. This observation has been appreciated
since Tamm's pioneering work on the electromagnetics of curved
spacetime in the 1920s \c{Skrotskii,Plebanski}, and it also provides
the foundation for the well-established C method in grating
analysis, developed by Chandezon and colleagues
\c{Chandezon_JOSA,Chandezon_AO} and applied in the context of
negative refraction \c{DIL2006}. Additionally, it  underpins recent
proposals of electromagnetic cloaks, which supposedly
 render objects contained inside the cloaks almost invisible to an external
 observer \c{Pendry}. These cloaks are made of composite materials with judiciously-engineered
microstructures or nanostructures.

A variety of theoretical schemes for achieving electromagnetic
cloaking have been put forward
\c{Milton_OE,Greenleaf,Lai,Leonhardt_NJP}. Furthermore, partial
cloaking in the microwave frequency range~---~at a single frequency
for a single polarization~---~was observed in the laboratory five
years ago for a cloak made from an assembly of split-ring resonators
\c{Schurig}. More recently, there have been further experimental
reports of cloaking involving tapered waveguides \c{Shalaev} and
plasmonic composite materials \c{Engheta}, for examples.

Designers of composite materials face severe challenges when
attempting to realize a cloak. Typically, a high degree of
anisotropy and a high degree of inhomogeneity are needed. In order
to cope with these requirements, cloak designs are generally based
on rather complex microstructures or nanostructures. However, some
simplifications have emerged very recently: A lamination process can
be implemented in order to realize a wide range of dielectric and
magnetic constitutive parameter values, such as may be required for
an electromagnetic cloak \c{Milton_NJP}.  Also, mention should be
made of a relatively simple cloak design based on an array of
elliptical rods proposed by Gao \emph{et al.} \c{Gao_OC}.

The aim of realizing a cloak using a simple microstructure or
nanostructure provides the motivation for our study. In the
following we show how the homogenization of a random distribution of
oriented spheroidal particles,  made from isotropic
dielectric-magnetic materials, can give rise to a well-functioning quasi-monochromatic cloak,
as demonstrated by a study of ray trajectories. In addition, we show
how the constitutive and morphological parameters of the component
materials, as well as their volume fractions, can be estimated using
a process of inverse homogenization.

In the following,  vectors are underlined, with the $\hat{}$ symbol
signifying a unit vector.   With respect to the Cartesian unit basis
vectors $\lec \hat{\#x}, \hat{\#y}, \hat{\#z} \ric$, the position
vector is expressed as  $\#r = x \, \hat{\#x}+ y \, \hat{\#y}+ z \,
\hat{\#z} $. Dyadics are double underlined. The permittivity and
permeability of free space in the absence of a gravitational field
are denoted as $\epso$ and $\muo$, respectively.

\section{Cloak inspired by a cosmic string}

\begin{figure}[!h]
\centering \psfull \epsfig{file=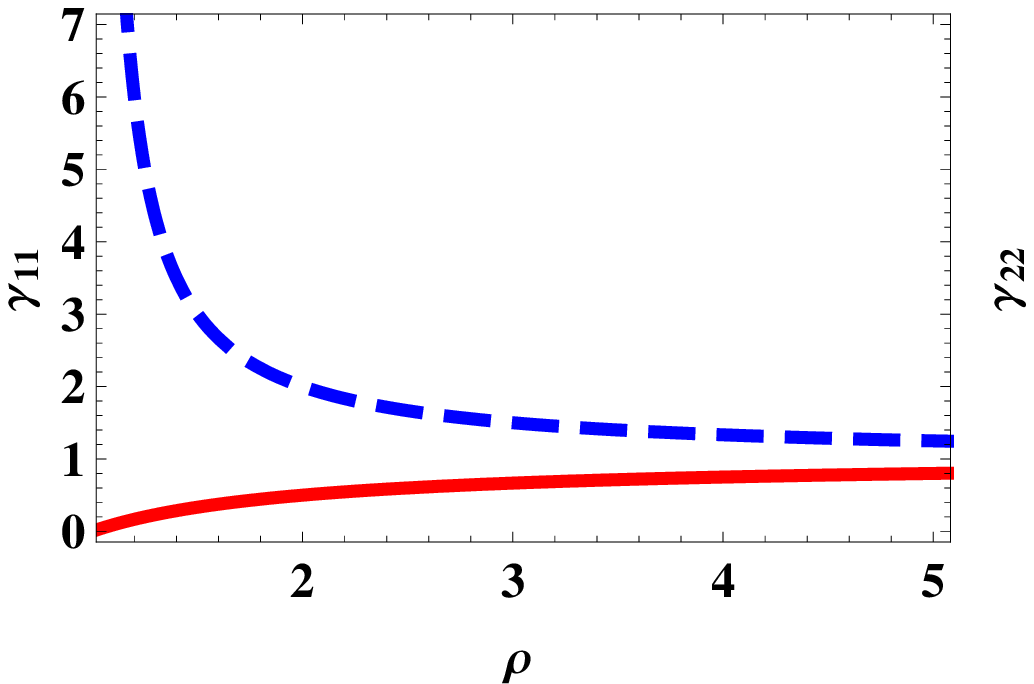,width=3.5in}
 \caption{The constitutive parameters $\gamma_{11}$  (red, solid curve) and $\gamma_{22}$ (blue, dashed curve) plotted versus
radial distance $\rho $. }\label{fig1}
\end{figure}

The object to be cloaked is contained within an empty cylinder which
is aligned with the $z$ axis and has radius $\rho = \rho_c > 1$,
where the radial distance $\rho = \sqrt{x^2 + y^2}$. The
constitutive relations of the inhomogeneous material from which the
cloak is composed are
\begin{equation}
\left. \begin{array}{l} \#D (\#r) = \epso \=\gamma (\rho) \. \#E
(\#r) \vspace{4pt}
 \\
\#B (\#r) = \muo \=\gamma (\rho) \. \#H (\#r)
\end{array}
\right\}, \qquad \qquad  \rho > \rho_c,
 \l{CR}
\end{equation}
where  the diagonal 3$\times$3 dyadic
\begin{equation}
\=\gamma (\rho) = \gamma_{11} (\rho) \le \hat{\#x}\,\hat{\#x} +
\hat{\#z}\,\hat{\#z} \ri + \gamma_{22} (\rho) \hat{\#y}\,\hat{\#y}
 \l{matrix_gamma}
\end{equation}
has  components
\begin{equation}
\l{g11g22}
\left.
\begin{array}{l}
\displaystyle{\gamma_{11} (\rho) = \frac{\rho - 1}{\rho}} \vspace{6pt} \\
\displaystyle{\gamma_{22} (\rho) = \frac{\rho}{ \rho - 1}} \\
\end{array}
\right\}. \end{equation}
 The constitutive parameters $\gamma_{11}  $ and $\gamma_{22} $ are plotted versus radial distance $\rho $ in
Fig.~\ref{fig1}. Both constitutive dyadics are obviously
positive definite for all $\rho > 1$. In the limit $\rho \to 1$, we
see that $\gamma_{11}$ becomes null-valued while $\gamma_{22}$
becomes unbounded. On the other hand, both $\gamma_{11}$ and
$\gamma_{22}$ rapidly approach unity as $\rho\to\infty$.

In fact, the structure of our proposed cloak is identical to  the
Tamm medium which represents the exterior region of a (non-spinning)
cosmic string \c{ML_String}. In a geometric-optics study of this
Tamm medium, it was demonstrated that the interior of the cosmic
string is almost entirely invisible to a distant observer
\c{AML_String}. Furthermore, the similarity between the constitutive
relations \r{CR}--\r{g11g22} and those of the
experimentally-realized cloak of Schurig \emph{et al.} \c{Schurig}
is noted.

\section{Ray trajectories}

A study of light rays passing through the cloak specified by the
constitutive relations \r{CR}--\r{g11g22} was performed, within the
geometric optics regime. As our ray-tracing methodology is described
in detail elsewhere \c{AML_String,AML_Alcubierre},  only a brief
outline is provided here. For a quasi-planewave with relative
wavevector $\#k$, the quantity
\begin{equation}
\mathcal{H}  =  \det \=\gamma (\rho) - \#k   \cdot \=\gamma (\rho )
\cdot
 \#k   \l{disp}
\end{equation}
represents a suitable Hamiltonian function. With the
parameterizations $\#r (\tau)$ and $\#k(\tau)$, ray trajectories are
governed by the coupled vector differential equations\footnote{Here
the shorthand $\nabla_{\#q} \equiv\hat{\#x}\,
\partial/ \partial q_x + \hat{\#y} \,\partial/ \partial q_y+\hat{\#z}\,\partial/ \partial
q_z $ for $\#q =  q_x\,\hat{\#x}+q_y\,\hat{\#y}+q_z \,\hat{\#z}$ is
adopted.} \c{Kline, Sluij2}
\begin{equation}
\left. \begin{array}{l} \displaystyle{
 \frac{d \#r}{d \tau} =  \nabla_{\#k}
 \mathcal{H}} \vspace{10pt} \\
\displaystyle{\frac{d \#k}{d \tau} = -  \nabla_{\#r}
 \mathcal{H}}
\end{array} \right\}, \l{odes}
\end{equation}
with their direction being given by $ \nabla_{\#k} \mathcal{H} $,
which is aligned with the time-averaged Poynting vector. Standard
numerical methods, such as the  Runge--Kutta method, can be used to
solve  the system \r{odes}  for $\#r( \tau)$ and $\#k (\tau)$, once
  appropriate initial conditions $\#r(0)$ and $\#k(0)$ have been
  specified.

\begin{figure}[!h]
\centering \psfull
\epsfig{file=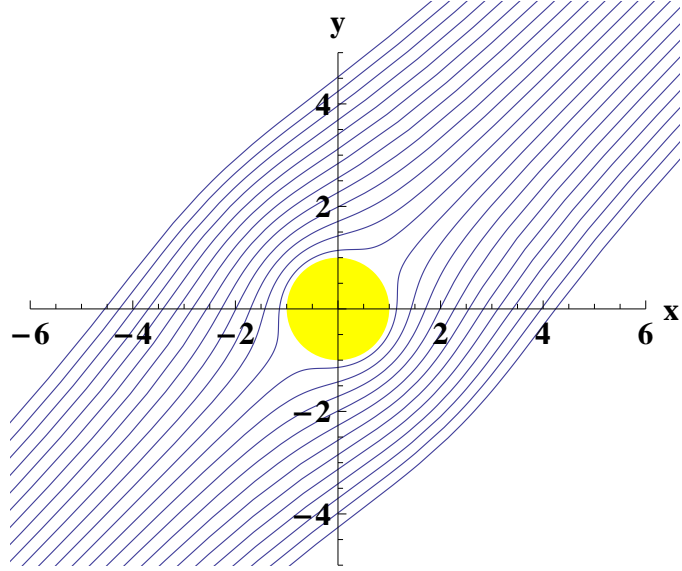,width=3.5in}
 \caption{A two-dimensional  example of
ray trajectories for $\rho_c = 1.1$.  Rays start at equally spaced
locations along the line $\#x (0) = \underline{\underline{M}} \. \le
30, \nu, 0 \ri $ with $-12.5 < \nu < 12.5 $ and where the rotation
dyadic $\underline{\underline{M}} = \cos \varphi \, \le \hat{\#x} \,
\hat{\#x} +  \hat{\#y} \, \hat{\#y} \ri - \sin \varphi \, \le
\hat{\#x} \, \hat{\#y} -  \hat{\#y} \, \hat{\#x} \ri $ with $\varphi
= \pi/4$; and $\#k (0)$ is directed along $  \le -1, -1, 0 \ri $.
}\label{fig2}
\end{figure}

An example of planar  ray trajectories is illustrated in
Fig.~\ref{fig2} for $\rho_c = 1.1$. Yellow shading indicates the
disc $\rho = 1$. The rays start at equally-spaced locations in the
first quadrant of $xy$ plane along a line parallel to $y=-x$, with
$\#k (0)$ directed along $\le -1, -1, 0 \ri$. The ray trajectories
remain in the $xy$ plane. The rays skirt around the outside of the
circle $\rho = \rho_c$, such that within a few radiuses from the
centre of the circle there is minimal deflection of the rays from
their original straight line trajectories. Thus the contents of the
cloak are almost entirely invisible to a distant observer located in
the third quadrant of the $xy$ plane.

\begin{figure}[!h]
\centering \psfull  \epsfig{file=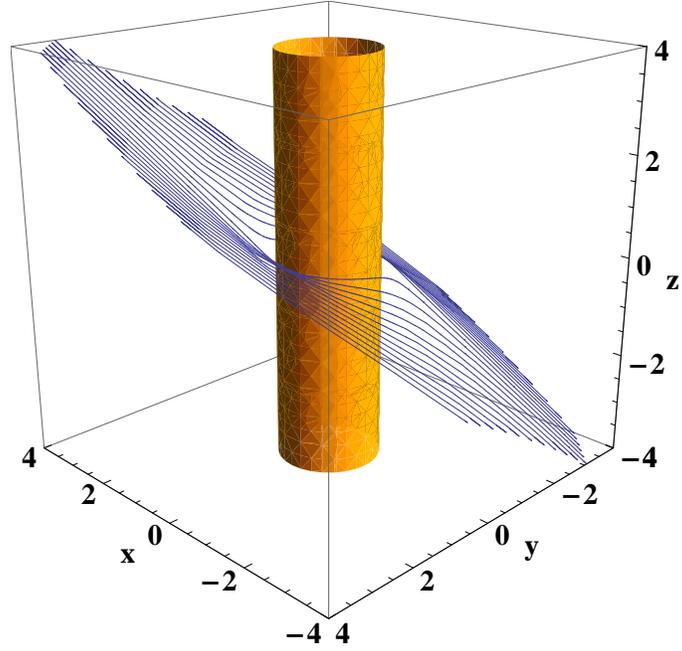,width=3.5in}
 \caption{A three-dimensional example of
ray trajectories for $\rho_c = 1.1$.  Rays start at equally spaced
locations along the line $\#x (0) = \underline{\underline{M}} \. \le
30, \nu, 29 \ri $ with $-12.5 < \nu < 12.5 $ and where the rotation
dyadic $\underline{\underline{M}}$ is defined as in Fig.~\ref{fig2};
and $\#k (0)$ is directed along $ \le -1, -1, -1 \ri$. }\label{fig3}
\end{figure}

An example of ray trajectories in three dimensions is represented in
Fig.~\ref{fig3}. Here  rays start at equally-spaced locations within
a plane parallel to $y=-x$ in the octant $\lec x>0, y>0, z>0\ric$,
with $\#k (0)$ directed along $\le -1, -1, -1 \ri$. The rays do not
remain in a single plane in this case, but the deviation from the
plane is relatively small, especially at distances beyond a few
radiuses from the cloak's axis.

By considering values of $\rho_c$ smaller than $1.1$, the degree of
cloaking is   improved  very little as perceived by an observer
more than a few radiuses away from the cloak's axis. However, by
increasing the value of $\rho_c$ from $1.1$ the degree of cloaking
quite rapidly diminishes.

\section{Cloak as a homogenized composite material}

Let us now turn to the issue of realizing the cloak specified by the
constitutive relations \r{CR}--\r{g11g22}. It was recently   demonstrated that
certain uniaxial dielectric-magnetic Tamm mediums can be envisaged
as homogenized composite materials (HCMs) that can be fabricated from relatively
simple component materials \c{ML_PRB}. A similar approach is taken
here.

Before turning to the homogenization process itself, we must first
deal with the inhomogeneity of the proposed cloak. This may be
catered for by subdividing the region of space occupied by the cloak
into local neighbourhoods which are sufficiently small as to be
considered approximately homogeneous. Then the inverse
homogenization procedure described later   in this  section can be
applied locally. This piecewise  homogeneous approach is documented
in detail elsewhere for general Tamm mediums \c{MLS_NJP}.

Now we turn to the homogenization of simple-component composite materials,
with our aim being  to specify a HCM whose constitutive parameters
coincide with the proposed cloak. As our theoretical foundation, we
rely on the well-established Bruggeman formalism \c{EAB,M_JNP}.

Suppose we consider the homogenization of a composite material comprising two isotropic component
materials,
 one being a  dielectric-magnetic material with
relative permittivity $\eps_a$ and relative permeability $\mu_a$ and
the other likewise being a dielectric-magnetic material but with
relative permittivity $\eps_b$ and relative permeability $\mu_b$.
The  component materials $a$ and $b$  are  randomly distributed with
respective volume fractions $f_a$ and $f_b = 1- f_a$. Each component
material is distributed as  spheroidal particles. The axis of these
spheroids for both component materials is assumed to be aligned with
the symmetry axis of $\=\gamma$, namely the $y$ axis. Thus, the
vector
\begin{equation}
\#r_{\,s} = \nu_\ell \, \le  \hat{\#x}\, \hat{\#x} + U_\ell \,
\hat{\#y}\, \hat{\#y} + \hat{\#z}\, \hat{\#z}   \ri \cdot \hat{\#r},
\qquad \quad \le \ell = a, b \ri,
\end{equation}
prescribes
 the surface of each spheroid relative to its
centre. Here the  vector $\hat{\#r}$ prescribes the surface of the
unit sphere, while the parameter $\nu_\ell > 0$  is a linear measure
of size which is taken to be much smaller than the electromagnetic
wavelengths involved.
 Prolate spheroids are characterized by $U_\ell > 1$
whereas  oblate spheroids are characterized by $U_\ell < 1$.

In accordance with the Bruggeman formalism, the corresponding HCM is
a uniaxial dielectric-magnetic material. Its relative permittivity
dyadic $\=\eps^{Br}$
 and relative
permeability dyadic $\=\mu^{Br}$ have the form
\begin{equation}
\=\tau =  \tau^{Br}_{11} \le \hat{\#x}\,\hat{\#x} +
\hat{\#z}\,\hat{\#z} \ri + \tau^{Br}_{22} \hat{\#y}\,\hat{\#y},
\qquad \le \tau = \eps, \mu \ri.
\end{equation}
For the  case of  uniaxial dielectric-magnetic HCMs such as is
involved here, full details of the numerical process of computing
the dyadics $\=\eps^{Br}$ and $\=\mu^{Br}$ are available elsewhere
\c{ML_PRB,EAB}.

Usually,  homogenization formalisms are implemented in the forward
sense, whereby the constitutive parameters of the HCM are estimated
from a knowledge of the constitutive and morphological parameters of
their component materials. However, our goal here is to estimate
constitutive and morphological parameters of the component materials
which  give rise to a HCM such that $\eps^{Br}_{11} = \mu^{Br}_{11}
= \gamma_{11}$ and $\eps^{Br}_{22} = \mu^{Br}_{22} = \gamma_{22}$.
In order to do so, an inverse implementation of the
 Bruggeman formalism is required. We note that formal expressions
of the inverse Bruggeman formalism have been established
\c{WSW_MOTL}, but in some instances these formal expressions can be
ill-defined \c{Cherkaev}. In practice, direct numerical methods may
be more effective in  implementing the inverse formalism \c{ML_JNP}.
A note of caution should be added concerning the inverse Bruggeman
formalism: certain constitutive parameter regimes
 have been identified as problematic \c{SSJ_TGM}, but these  regimes
 do not overlap with
 those
 considered here.

As a representative example, let us focus on a particular inverse
homogenization scenario whilst noting that other scenarios are also
feasible. For convenience, we seek component materials which are
composed of spheroidal particles all of the same shape; i.e., $U_a
\equiv U_b$. Furthermore, we shall assume that the relative
permittivity and relative permeability parameters of the component
materials are the same; i.e., $\eps_a \equiv \eps_b$ and $\mu_a
\equiv \mu_b$. This coincidence of constitutive parameters can be
straightforwardly engineered if the component materials themselves
are envisaged as HCMs, as has been demonstrated previously
\c{ML_PRB}. Now, assuming that the relative permittivity and
relative permeability of component material $a$ are fixed, we
implement the inverse Bruggeman formalism in order to estimate the
 volume faction, shape parameter and  constitutive parameters for
component material $b$.

 The inversion of the Bruggeman formalism may carried out as
 follows.
Let  $\lec \breve{\eps}^{Br}_{11}, \breve{\eps}^{Br}_{22},
\breve{\mu}^{Br}_{11}, \breve{\mu}^{Br}_{22} \ric $ denote the
forward Bruggeman estimates of the HCM's relative permittivity and
relative permeability parameters, computed  for physically-plausible
ranges of the parameters $f_b$, $U_{b}$ and $\eps_b$,
 namely $ f_b  \in
\le f^-_b, f^+_b \ri$,  $U_{b} \in \le U^-_{b}, U^+_{b} \ri$ and
$\eps_{b} \in \le \eps^-_{b}, \eps^+_{b} \ri$. Then:
\begin{itemize}
\item[(i)] Fix $ U_b = \le U^-_b + U^+_b \ri /2$
and $ \eps_b = \le \eps^-_b + \eps^+_b \ri /2$. For all values of
$f_b \in \le f^-_b, f^+_b \ri$, find the value $f^\dagger_b$  for
which the quantity
 \begin{eqnarray} \Delta &=& \les \le
\frac{\breve{\eps}^{Br}_{11}  - \gamma_{11}}{\gamma_{11}} \ri^2 +
\le \frac{\breve{\eps}^{Br}_{22}  - \gamma_{22}}{\gamma_{22}} \ri^2
 + \le \frac{\breve{\mu}^{Br}_{11}  - \gamma_{11}}{\gamma_{11}}
\ri^2  + \le \frac{\breve{\mu}^{Br}_{22} - \gamma_{22}}{\gamma_{22}}
\ri^2 \ris^{1/2}
\end{eqnarray} is minimized.
\item[(ii)] Fix $f_b = f^\dagger_b$ and $ \eps_b = \le \eps^-_b + \eps^+_b \ri /2$.
 For all values of $U_b \in \le U_b^-, U_b^+ \ri$, find the
value $U_b^\dagger$  for which $\Delta$ is minimized.
\item[(iii)] Fix $f_b = f^\dagger_b$ and $U_b = U^\dagger_b$.
 For all values of $\eps_b \in \le \eps_b^-, \eps_b^+ \ri$, find the value
$\eps_b^\dagger$ for which  $\Delta$ is minimized.
\end{itemize}
The steps (i)--(iii) are repeated, using $U_b^\dagger$ and
$\eps_b^\dagger$ as the fixed values of $U_{b}$ and $\eps_b$
 in step (i), and  $\eps_b^\dagger$ as the fixed value of
 $\eps_b$ in step (ii),
 until $\Delta$ becomes sufficiently
small.

\begin{figure}[!h]
\centering \psfull \epsfig{file=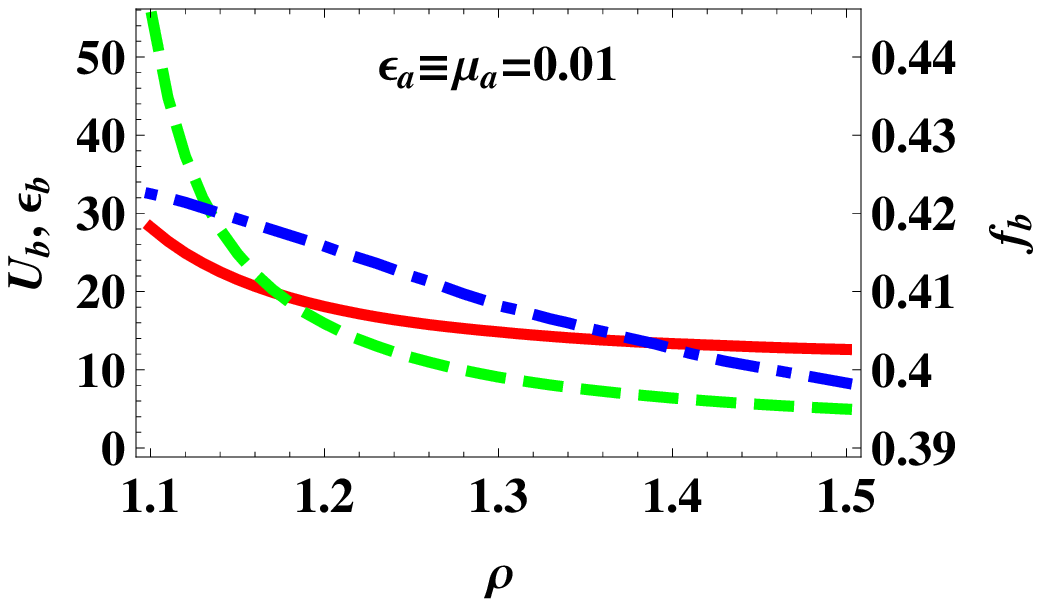,width=3.5in}
\epsfig{file=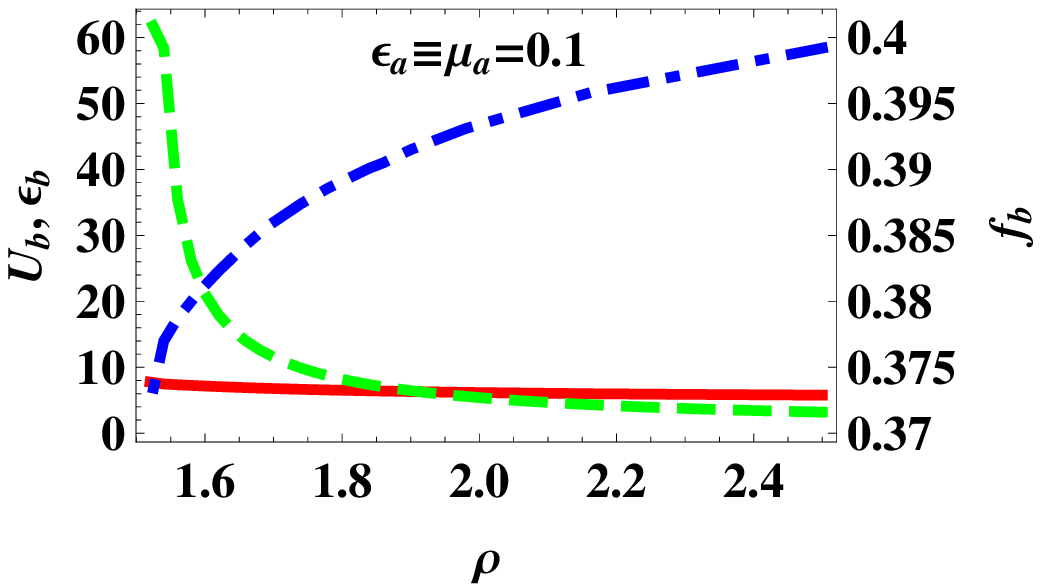,width=3.5in}
\epsfig{file=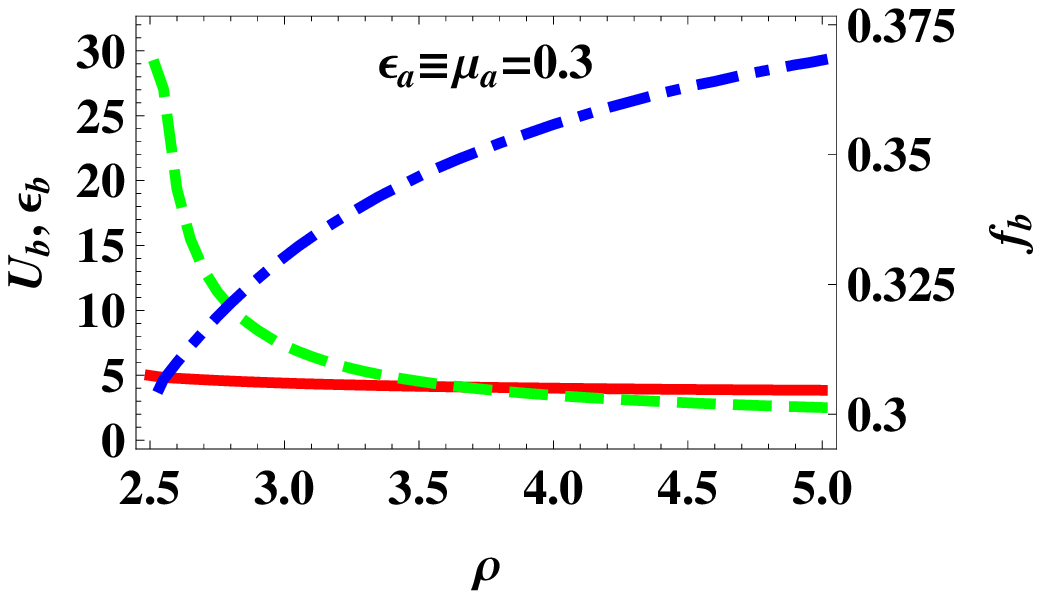,width=3.5in}
 \caption{The constitutive parameter $\eps_b \equiv \mu_b$ (red, solid curves), shape parameter $U_b \equiv
 U_a$ (green, dashed curves), and the volume faction $f_b$ (blue, broken dashed curves) plotted versus radial distance
 $\rho$ for (a) $\eps_a \equiv \mu_a = 0.01$, (b) $\eps_a \equiv \mu_a =
 0.1$, and (c)  $\eps_a \equiv \mu_a = 0.3$.
} \label{fig4}
\end{figure}

Estimates of $f_b$, $U_b$ and $\eps_b$ computed using this scheme
are presented in Fig.~\ref{fig4} for the case $\rho_c = 1.1$. These
quantities are plotted versus the radial distance $\rho$. The
relative permittivities  chosen for component material $a$ were:
$\eps_a = 0.01$ for the range $\rho_c < \rho \leq 1.5$; $\eps_a =
0.1$ for the range $1.5 < \rho \leq 2.5$; and $\eps_a = 0.3$ for the
range $2.5 < \rho \leq 5$. For all the  numerical results presented
in
 Figs.~\ref{fig4}, the degree of
 convergence of the numerical schemes
 was $< 0.1 \%$. While the constitutive parameters chosen for
 component material $a$ are not commonplace, reports of materials with relative
permittivities and relative permeabilities close to zero are to be
found in the recent literature
 \c{Alu,Lovat,Cia_PRB}. Materials with extreme properties are now being considered as well
 \c{STdB2007}.
 Furthermore, the constitutive parameters for
 component material $b$ represented in Fig.~\ref{fig4} are not at all
 exotic.

\section{Closing remarks}

We have shown theoretically that the homogenization of a composite
material comprising two simple isotropic component materials can
result in an HCM which can be used to create a well-functioning
cylindrical cloak. By carefully controlling the spheroidal shape of
the particles which make up the two component materials, the high
degree of HCM anisotropy which is essential for cloaking can be
achieved. The inverse Bruggeman formalism provides estimates of the
 shapes  needed for the component particles, as well as   their
 appropriate
constitutive parameters and volume fractions. Provided the component materials
are weakly dispersive in their constitutive parameters, the proposed cloak can function
in a spectral regime of appreciable bandwidth.

The homogenization scenario presented here is by no means the only
way of realizing a uniaxial dielectric-magnetic HCM of the required
form. For example, the two component materials could  both be
uniaxial dielectric-magnetic materials (with parallel symmetry axes)
which are distributed as spherical particles \c{MW_JOPA}.
Alternatively,
 four component materials could be homogenized: two isotropic dielectric
materials and two isotropic magnetic materials, with all four
component materials being composed of spheroidal particles (all with
the same alignment) \c{ML_PRB}.
 However, the
two-component formulation presented here is more
computationally-efficient for the inverse Bruggeman formalism.

\vspace{10mm}

\noindent {\bf Acknowledgment:}  AL thanks the Charles Godfrey
Binder Endowment at Penn State for partial financial support of his
research activities.


\newpage

\newpage

\newpage

\newpage

\end{document}